\documentclass[11pt]{amsart}

\copyrightinfo{}{}

\newtheorem{theorem}{Theorem}[section]
\newtheorem{lemma}[theorem]{Lemma}

\theoremstyle{definition}
\newtheorem{definition}[theorem]{Definition}

\theoremstyle{remark}

\numberwithin{equation}{section}

\newcommand{\abs}[1]{\lvert#1\rvert}

\newcommand{\BB}[1]{\ensuremath{\mathbb{#1}}}
\newcommand{\R}{\ensuremath{\BB{R}}}
\newcommand{\iny}{\ensuremath{\infty}}
\newcommand{\grad}{\ensuremath{\nabla}}
\DeclareMathOperator{\dv}{div}
\newcommand{\prt}{\ensuremath{\partial}}

\newcommand{\pr}[1]{\ensuremath{\left( #1 \right) }}
\newcommand{\set}[1]{\ensuremath{\left\{ #1 \right\}}}
\newcommand{\norm}[1]{\ensuremath{\left\Vert #1 \right\Vert}}
\newcommand{\smallnorm}[1]{\ensuremath{\Vert #1 \Vert}}
\newcommand{\refS}[1]{Section~\ref{S:#1}}
\newcommand{\refT}[1]{Theorem~\ref{T:#1}}
\newcommand{\refL}[1]{Lemma~\ref{L:#1}}
\newcommand{\refD}[1]{Definition~\ref{D:#1}}

\newcommand{\refE}[1]{Equation~(\ref{e:#1})}
\newcommand{\eps}{\ensuremath{\epsilon}}
\newcommand{\Cal}[1]{\ensuremath{\mathcal{#1}}}

\newcommand{\myabs}[1]{\ensuremath{\left\vert #1 \right\vert}}

\newcommand{\LOneNorm}[1]
    {\norm{#1}_{L^1}}

\newcommand{\LTwoNorm}[1]
    {\norm{#1}_{L^2}}

\newcommand{\LpNorm}[2]
    {\norm{#1}_{L^{#2}}}

\newcommand{\SmallLpNorm}[2]
    {\smallnorm{#1}_{L^{#2}}}

\begin{document}

\title
    [Inviscid limit with unbounded vorticity]
    {The inviscid limit for two-dimensional incompressible fluids
        with unbounded vorticity}

%--- Information for first author
\author{James P. Kelliher}
% --- Address of record for the research reported here
\address{Department of Mathematics, University of Texas, Austin,
         Texas, 78712}
%--- Current address
\curraddr{Department of Mathematics, University of Texas, Austin,
          Texas, 78712}
\email{kelliher@math.utexas.edu}
% \thanks{}

%    General info
 % ; Secondary }
\date{June 26, 2003} %  and, in revised form, .}

% \dedicatory{}


\begin{abstract}
    In \cite{C1996}, Chemin shows that solutions of the Navier-Stokes
    equations in $\R^2$ for an incompressible fluid whose initial vorticity
    lies in $L^2 \cap L^\iny$ converge in the zero-viscosity limit in the
    $L^2$--norm to a solution of the Euler equations, convergence being
    uniform over any finite time interval. In \cite{Y1995}, Yudovich
    assumes an initial vorticity lying in $L^p$ for all $p \ge p_0$, and
    establishes the uniqueness of solutions to the Euler equations for an
    incompressible fluid in a bounded domain of $\R^n$, assuming a
    particular bound on the growth of the $L^p$--norm of the initial
    vorticity as $p$ grows large. We combine these two approaches to
    establish, in $\R^2$, the uniqueness of solutions to the Euler
    equations and the same zero-viscosity convergence as Chemin, but under
    Yudovich's assumptions on the vorticity with $p_0 = 2$. The resulting
    bounded rate of convergence can be arbitrarily slow as a function of
    the viscosity $\nu$.
\end{abstract}

\maketitle

\section{Introduction}\label{S:Introduction}
The equations of motion governing an incompressible fluid with
viscosity $\nu$ are the Navier-Stokes equations,
\begin{align*}
    \begin{matrix}
        (NS_\nu) & \left\{
            \begin{matrix}
                \prt_t v_\nu + v_\nu \cdot \grad v_\nu - \nu \Delta
                    v_\nu = - \grad p_\nu \\
                \dv v_\nu = 0 \\
                v_\nu|_{t = 0} = v^0.
            \end{matrix}
            \right.
    \end{matrix}
\end{align*}
These same equations with zero viscosity become the Euler equations:
\begin{align*}
    \begin{matrix}
        (E) & \left\{
            \begin{matrix}
                \prt_t v + v \cdot \grad v = - \grad p\\
                \dv v = 0 \\
                v|_{t = 0} = v^0.
            \end{matrix}
            \right.
    \end{matrix}
\end{align*}

The question of whether a solution to ($NS_\nu$) converges, by some
measure, to a solution to ($E$) as $\nu \to 0$ (the \textit{inviscid}
or \textit{zero-viscosity} limit) has a long history. Temam has a
discussion of this in Appendix III of \cite{T2001}. See also Kato's
remarks in \cite{K1983}. Briefly, convergence of smooth solutions in
$\R^n$ is well understood. Much less is known about convergence of weak
solutions in $\R^n$ or the convergence of solutions, weak or smooth, in
a domain with boundaries.

We restrict our attention to fluids extending throughout $\R^2$, with
the initial velocity belonging, for some real number $m$, to the space
$E_m$ of \cite{C1996} and \cite{C1998}. A vector $v$ belongs to $E_m$
if it is divergence-free and can be written in the form $v = \sigma +
v'$, where $v'$ is in $L^2(\R^2)$ and where $\sigma$ is a
\textit{stationary vector field}, meaning that $\sigma$ is of the form,
\begin{align}\label{e:Stationary}
    \sigma = \pr{-\frac{x^2}{r^2} \int_0^r \rho g(\rho) \, d \rho,
                \;\frac{x^1}{r^2} \int_0^r \rho g(\rho) \, d \rho},
\end{align}
where $g$ is in $C_0^\iny(\R \setminus \set{0})$. $E_m$ is an
affine space; fixing an origin, $\sigma$, in $E_m$ we can define a
norm by $\norm{\sigma + v'}_{E_m} = \LTwoNorm{v'}$. Convergence in
$E_m$ is equivalent to convergence in the $L^2$--norm to a vector
in $E_m$.

We use the notation $\omega(v)$, or just $\omega$ when $v$ is
understood, for the vorticity of $v$, which equals $\prt_1 v^2 - \prt_2
v^1$. The initial vorticity we denote by $\omega^0$.

The following is a fundamental result of Yudovich's (\cite{Y1963}), as
adapted by Chemin in \cite{C1991} from bounded domains to all of $\R^2$
(see \cite{C1998}):

\begin{theorem}[Yudovich's theorem] \label{T:YudovichsTheorem}
    Let $v^0$ be in $E_m$, with $\omega^0$ belonging to $L^a(\R^2)
    \cap L^\iny(\R^2)$ for some $1 < a < \iny$. Then there
    exists a unique solution $v$ of ($E$) belonging to $C(\R; E_m)$
    such that $\omega(v)$ is in $L^\iny(\R^3) \cap L^\iny(\R; L^a(\R^2))$.
\end{theorem}

%
%------ Revision: 24 April 2004: Reference to Serfati added.
%
In \cite{Y1995}, Yudovich, in the setting of a bounded domain in
$\R^n$ with impermeable boundary, weakens the conditions on the
initial vorticity in \refT{YudovichsTheorem}, allowing unbounded
vorticity, and is still able to obtain uniqueness. (Similar
results have been obtained by Serfati in \cite{S1994}.) Chemin
shows in \cite{C1996} that with the assumptions on the initial
data in \refT{YudovichsTheorem} with $a = 2$, solutions
$(v_\nu)_{v > 0}$ of ($NS_\nu$) converge in the $L^2$--norm
uniformly over a finite time interval as $\nu \to 0$ to the unique
solution $v$ of ($E$) given by \refT{YudovichsTheorem}. We
establish the same convergence as Chemin, but with the initial
vorticity of Yudovich.

To describe Yudovich's conditions on the initial vorticity, let
$\phi \colon [p_0, \iny) \to \R^+$ be a continuous function, where
$p_0 > 1$. We define two functions, $\beta_{\eps, M, \phi} \colon
\R^+ \to \R^+$ and $\beta_{M, \phi}: \R^+ \to \R^+$, parameterized
by $\eps$ in $(0, 1/p_0]$, $M > 0$, and $\phi$:
\begin{align}\label{e:BetaExpression}
    \begin{split}
        &\beta_{\eps, M, \phi}(x)
            = M^\eps x^{1 - \eps} \phi(1/\eps), \\
        &\beta_{M, \phi}(x)
            = \inf \set{\beta_\eps(x): \eps \in (0, 1/p_0]}.
    \end{split}
\end{align}
For brevity, we usually write $\beta_\eps$ for $\beta_{\eps, M, \phi}$
and $\beta$ for $\beta_{M, \phi}$, with the choices of $M$ and $\phi$
being understood.

For all $\eps$ in $(0, 1/p_0]$, $\beta_\eps(x)$ is a monotonically
increasing function continuous in $x$ and in $\eps$, with $\lim_{x \to
0^+} \beta_\eps(x) = 0$. It follows that $\beta$ is a monotonically
increasing continuous function and that $\lim_{x \to 0^+} \beta(x) =
0$. Also, $\beta(x) \le \beta_\eps(x)$ for all $\eps$ in $(0, 1/p_0]$
and $x \in \R^+$.

\begin{definition}\label{D:Admissible}
    A continuous function $\theta \colon [p_0, \iny) \to \R^+$ is called
    \textit{admissible} if
    \[
        \int_{0}^1 \frac{ds}{\beta_{M, \phi}(s)} = \iny,
    \]
    where $\phi(p) = p \theta(p)$. This condition is independent of the
    choice of $M$.
\end{definition}

%
%------ Revision: Next paragraph inserted 24 April 2004
%
In \refS{YudovichBounds} we give examples of admissible functions
and discuss how our definition relates to the equivalent
definition in \cite{Y1995}.

Yudovich proves that for a bounded domain in $\R^n$ with
impermeable boundary (which adds the condition to ($E$) that the
normal component of the velocity on the boundary is zero), if the
$L^p$--norms of the initial vorticity are bounded by an admissible
function $\theta$, then at most one solution to ($E$) exists.

For our purposes, we define (weak) solutions to ($E$) and ($NS_\nu$) as
follows:

\begin{definition}\label{D:WeakSolution}
    A time-varying vector field $v \colon \R \times \R^2 \to \R^2$ is a weak
    solution to ($E$) or ($NS_\nu$) if there exists a distribution $p$ such
    that ($E$) or ($NS_\nu$) hold in the sense of distributions and if, in
    addition,
    \begin{itemize}
        \item[(i)]
            $v$ is in $L^\iny_{loc}(\R; E_m)$ for some real $m$, and
        \item[(ii)]
            there exists a $p_0 > 1$ such that $\grad v$
            is in $L^\iny_{loc}(\R; L^p(\R^2))$ for all $p$ in $[p_0, \iny)$.
    \end{itemize}
\end{definition}

We combine the techniques of Chemin and Yudovich to prove the following
theorem:
\begin{theorem}\label{T:MainResult}
    Let $v^0$ be in $E_m$ and assume that $\omega^0$ is in $L^p(\R^2)$
    for all $p$ in $[2, \iny)$,
    with $\SmallLpNorm{\omega^0}{p} \le \theta(p)$ for
    some admissible function $\theta$. Then:
    \begin{itemize}
        \item[(i)]
            There exists a unique solution $v$ of ($E$).
        \item[(ii)]
            For all $\nu > 0$, there exists a unique solution $v_\nu$ of
            ($NS_\nu$).
        \item[(iii)]
            $\LTwoNorm{v_\nu(t) - v(t)} \to 0$ in $L^2(\R^2)$ uniformly on $[0,
            T]$ as $\nu \to 0^+$.
    \end{itemize}
\end{theorem}

We prove only the uniqueness statements of (i) and (ii), a proof
of existence following from the bounds we obtain on the
$L^2$--norm of the difference between two solutions, much as in
the proof of \refT{YudovichsTheorem}. It is also true that $v$ and
$v_\nu$ lie in $C(\R; E_m)$ (after possibly changing their values
on a set of measure zero), but we do not use this fact.

Given an initial velocity in $E_m$, there exists a unique solution in
the sense of distributions to ($NS_\nu$) in $C([0, T]; E_m) \cap
L^2([0, T]; \dot{H}^1)$ for all $T > 0$. This is essentially a
classical result of Leray, which can be proved, for instance, by
straightforward modifications of the proofs of Theorems 3.1 and 3.2 of
Chapter 3 of \cite{T2001}. Additional assumptions, such as those of
\refT{MainResult}, are required, however, to conclude that the velocity
is in $L^\iny([0, T]; \dot{H}^1)$, not just in $L^2([0, T];
\dot{H}^1)$.

The rate of convergence in the inviscid limit is also of interest.
Constantin and Wu in \cite{CW1995} show that the $L^2$--rate of
convergence of the velocity for a vortex patch in $\R^2$ with
smooth boundary is $O(\sqrt{\nu t})$ uniformly over any finite
time interval, and remark that this same result holds when $\grad
v$ is in $L^1_{loc}(\R; L^\iny(\R^2))$, where $v$ is the solution
to ($E$). Chemin in \cite{C1996} gives essentially the same bound
on the convergence rate as that in \cite{CW1995}, assuming that
$v$ is in $L^\iny_{loc}(\R^+;Lip)$, which implies the condition in
\cite{CW1995} that $\grad v$ lie in $L^1_{loc}(\R; L^\iny(\R^2))$.

Chemin goes on to establish bounds on the rate of convergence given
initial vorticity in $L^2 \cap L^\iny$, the bounded rate of convergence
always being slower than $O(\sqrt{\nu})$, but approaching that order
for small time intervals. The approach we take leads, in the special
case of $L^2 \cap L^\iny$, to the same bound on the rate of convergence
as Chemin. In the general case of unbounded vorticity, however, the
bounded rate of convergence can be arbitrarily slow.

In \cite{CW1996}, Constantin and Wu consider an initial vorticity
in $\R^2$ lying in the space $\mathbf{Y}$ of bounded, compactly
supported functions. They also assume that the initial vorticity
lies in certain Besov spaces, and establish convergence of the
\textit{vorticity} in every $L^p$--norm for $p \ge 2$, with the
rate of convergence increasing with increasing $p$. In
\cite{CW1997}, the same authors consider statistical solutions of
($NS_\nu$) and their inviscid limits, working again with the space
$\mathbf{Y}$.

%
%------ Revision: Next paragraph inserted 24 April 2004
%
We also note that given the uniqueness of the solution to ($E$) in
$\R^2$ established in \refT{MainResult}, the compactness argument
on p. 131-133 of \cite{L1996} would imply the strong convergence
in (iii) of \refT{MainResult}. A bound on the rate of convergence
does not follow from that approach, however.

We use without proof the following theorem:

\begin{theorem}\label{T:FactsAboutSolution}
    Let $v$ be a solution to ($NS_\nu$) or ($E$) as defined in
    \refD{WeakSolution}, and let
    $\sigma$ be any stationary vector field in $E_m$. Then:
    \begin{itemize}
        \item[(i)]
            $v - \sigma$ is in $L_{loc}^\iny(\R; L^2(\R^2))$
            (\textit{i.e.}, the $L^2$--norm of $v - \sigma$ is bounded
            over any finite time interval), the norm being bounded
            over $\set{\nu > 0}$;
        \item[(ii)]
            $v$ is in $L_{loc}^\iny(\R; L^\iny(\R^2))$, the norm being bounded
            over $\set{\nu > 0}$;
        \item[(iii)]
            $\LpNorm{\omega(t)}{p} \le \SmallLpNorm{\omega^0}{p}$ for all $1 \le p
            \le \iny$;
        \item[(iv)]
            there exists a constant $C$ such that for all $p \ge 2$,
            $\LpNorm{\grad v}{p} \le Cp \LpNorm{\omega}{p}$ when $\omega$
            is in $L^p$.
    \end{itemize}
\end{theorem}

In \refT{FactsAboutSolution}, (i) comes from energy estimates, as does
(ii) after decomposing $v - \sigma$ into high and low frequencies.
Equality holds in (iii) for solutions to ($E$), and (iv) is a result
from harmonic analysis that applies to all divergence-free vector
fields in $\R^n$.

We will also need Osgood's lemma, the proof of which can be found, for
example, on p. 92 of \cite{C1998}.

\begin{lemma}[Osgood's lemma]\label{L:Osgood}
    Let $L$ be a measurable positive function and $\gamma$ a positive
    locally integrable function, each defined on the domain $[t_0,
    t_1]$. Let $\mu \colon [0, \iny) \to [0, \iny)$ be a continuous nondecreasing
    function, with $\mu(0) = 0$. Let $a \ge 0$, and assume that for
    all $t$ in $[t_0, t_1]$,
    \[
        L(t) \le a + \int_{t_0}^t \gamma(s) \mu(L(s)) \, ds.
    \]
    If $a > 0$, then
    \[
        -\Cal{M}(L(t)) + \Cal{M}(a)
            \le \int_{t_0}^t \gamma(s) \, ds,
        \;\text{where}\;
        \Cal{M}(x) = \int_x^1 \frac{ds}{\mu(s)}.
    \]
    If $a = 0$ and $\Cal{M}(0) = \iny$, then $L \equiv 0$.
\end{lemma}

%
%------ Revision: The following section added 24 April 2004
%

%
%------------- Yudovich's Unbounded Vorticity ------------------
%
\section{Yudovich's Unbounded Vorticity}\label{S:YudovichBounds}

\noindent \refD{Admissible} is equivalent to requiring that
\begin{align}\label{e:Psi}
    \psi(x) := \inf \set{(x^\eps/\eps) \theta(1/\eps):
                \eps \in (0, 1/p_0]}
\end{align}
satisfy
\[
    \int_1^\iny \frac{dx}{x \psi(x)} = \iny,
\]
which is essentially the same as the condition in \cite{Y1995}.
The functions $\psi$ and $\beta$ are related by $\psi(x) = x
\beta(1/x)$ when $M = 1$.

Choosing $\eps = 1/\ln x$ in \refE{Psi} shows that $\psi(x) \le e
(\ln x) \theta(\ln x)$ when $x \ge \exp(p_0)$. It follows that
\begin{align}\label{e:SimplerIntegralCondition}
    \int_1^\iny \frac{dx}{x \psi(x)}
        \ge \int_{e^{p_0}}^\iny
            \frac{dx}{e x (\ln x) \theta(\ln x)}
        = \frac{1}{e}
            \int_{p_0}^\iny \frac{dp}{p \theta(p)}.
\end{align}
For $\theta$ to be admissible, it is sufficient, though not
necessary, that the final integral in
\refE{SimplerIntegralCondition} be infinite. Thus we can say, as a
rough measure only, that the $L^p$--norm of the initial vorticity
can grow in $p$ only slightly faster than $\log p$ and still be
handled by our approach. Such growth in the $L^p$--norm arises,
for example, from a point singularity of the type $\log \log
(1/x)$.

Define, as in \cite{Y1995}, the sequence of admissible bounds on
vorticity,
\begin{align}\label{e:YudovichExamples}
    \theta_0(p) = 1,
    \theta_1(p) = \ln p, \dots,
    \theta_m(p) = \ln p \cdot \ln \ln p \cdots \ln^m p,
\end{align}
where $\ln^m$ is $\ln$ composed with itself $m$ times. These are
each admissible since $\psi(x) \le e (\ln x) \theta_m(\ln x) = e
\theta_{m + 1}(x)$, and a repeated change of variables shows that
the final integral in \refE{SimplerIntegralCondition} is infinite
for $\theta = \theta_m$.

%
%------------- Proof of Main Result ------------------
%
\section{Proof of \refT{MainResult}}\label{S:YudovicAndCheminMethods}

We take a unified approach to proving the three parts of
\refT{MainResult}. Let each of $(v_\nu)_{\nu > 0}$ and $(v'_\nu)_{\nu >
0}$ be either a family of solutions to ($NS_\nu$) parameterized by the
viscosity $\nu$ or a single solution to ($E$). In the latter case, the
solution is independent of the value of $\nu$. All solutions in
$(v_\nu)_{\nu > 0}$ and $(v'_\nu)_{\nu > 0}$ share the same initial
velocity $v^0$, which lies in $E_m$ and satisfies the vorticity bounds
assumed in \refT{MainResult}. Let
\[
    w_\nu = v_\nu - v'_\nu.
\]

\begin{theorem}\label{T:FromCheminApproach}
    Under the assumptions of \refT{MainResult}, for all $t \ge 0$,
    \begin{align*}
        \int_{\R^2} \abs{w_\nu(t, x)}^2 \, dx
            \le R \nu t + 2\int_0^t \int_{\R^2}
                \abs{\grad v'_\nu(s, x)} \abs{w_\nu(s, x)}^2 \, dx \, ds.
    \end{align*}
    $R = 0$ when $w_\nu$ is the difference between two solutions to
    ($NS_\nu$) and when $w_\nu$ is the difference between two solutions
    to ($E$). $R > 0$ when $w_\nu$ is the difference between a
    solution to ($NS_\nu$) and a solution to ($E$).
\end{theorem}
\begin{proof}
    See \refS{ProofOfThm23}.
\end{proof}

\begin{theorem}\label{T:YudovichTheorem}
    Let $f_\nu$ and $g_\nu$ be nonnegative measurable real
    functions on $[0, T] \times \R^2$ parameterized by
    $\nu > 0$ for some $T > 0$. Assume that $f_\nu(t)$ is in  $L^1(\R^2)$
    for all $t \in [0, T]$ and $\nu > 0$, and that
    \[
        \sup_{\nu > 0} \set{\norm{f_\nu}_{L^\iny([0, T] \times \R^2)}} < \iny.
    \]
    Assume that for some $p_0 > 1$ and some function $\phi$, where $\phi(p) =
    p \theta(p)$ for an admissible function $\theta$,
    \[
        \norm{g_\nu(t, \cdot)}_{L^p(\R^2)} \le \phi(p)
    \]
    for all $t$ in $[0, T]$ and $p \ge p_0$.
    Assume also that for some real constant $R$,
    \[
        L_\nu(t) :=
            \int_{\R^2} f_\nu(t, x) \, dx
                \le R \nu t + \int_0^t \int_{\R^2} g_\nu(s, x) f_\nu(s, x) \, dx \,
                    ds.
    \]

    If $R = 0$ then $L_\nu \equiv 0$.

    If $R > 0$ then $L_\nu(t) \to 0$ uniformly on $[0, T]$ as $\nu \to 0^+$.
\end{theorem}
\begin{proof}
    See \refS{ProofOfThm22}.
\end{proof}

\noindent \textbf{Proof of \refT{MainResult}}. Fix a $T > 0$ and let
$f_\nu = \abs{w_\nu}^2$, $g_\nu = 2\abs{\grad v'_\nu}$. Then
\[
    A = \sup_{\nu > 0} \set{\norm{v_\nu}_{L^\iny([0, T] \times \R^2)}}
    \text{ and }
    B = \sup_{\nu > 0} \set{\norm{v'_\nu}_{L^\iny([0, T] \times \R^2)}}
\]
are finite by \refT{FactsAboutSolution}, so
\begin{align*}
    \sup_{\nu > 0} \set{\norm{f_\nu}_{L^\iny([0, T] \times \R^2)}}
        \le (A + B)^2 < \iny.
\end{align*}

Also by \refT{FactsAboutSolution},
\[
    \norm{g_\nu}_{L^p(\R^2)}
        \le Cp \norm{\omega^0}_{L^p(\R^2)}
        \le \phi(p)
        = p (C \theta(p)),
\]
where $C \theta$ is an admissible function since $\theta$ is admissible
by assumption. Applying \refT{FromCheminApproach} and
\refT{YudovichTheorem} yields all three parts of \refT{MainResult}
(only the uniqueness portions of parts (i) and (ii), though; see the
comment following the statement of \refT{MainResult}). $\square$

%
%------------ Proof of a Theorem of Yudovich -----------------
%
\section{Proof of \refT{YudovichTheorem}}\label{S:ProofOfThm22}

In this section we prove \refT{YudovichTheorem} following the approach
in \cite{Y1995}.

\begin{lemma}\label{L:YudovichLemma}
    Let $D$ be a measurable subset of $\R^n$ and let $f$ and $g$ be nonnegative
    measurable real functions in $L^1(D) \cap L^\iny(D)$. Let $M \ge \norm{f}_{L^\iny(D)}$.
    Assume that for some $p_0 > 1$ and some positive function
    $\phi \colon [p_0, \iny) \to \R^+$,
    \[
        \norm{g}_{L^p(D)} \le \phi(p)
    \]
    for all $p \ge p_0$. Then
    \[
        \int_D f(x) g(x) \, dx
            \le \beta(\norm{f}_{L^1(D)}).
    \]
\end{lemma}
\begin{proof}
    Let $\eps$ be in $(0, 1/p_0]$. Then
    \begin{align*}
        \int_D f g
            &\le M^\eps \int_D f^{1 - \eps} g
            \le M^\eps \SmallLpNorm{f^{1-\eps}}{1/(1 - \eps)}
                \LpNorm{g}{1/\eps} \\
            &\le M^\eps \LOneNorm{f}^{1-\eps} \phi(1/\eps)
             = \beta_\eps(\LOneNorm{f}).
    \end{align*}
    The conclusion follows from \refE{BetaExpression}.
\end{proof}

\noindent \textbf{Proof of \refT{YudovichTheorem}}. Letting $M =
\sup_{\nu > 0} \set{\norm{f_\nu}_{L^\iny([0, T] \times \R^2)}}$ and $D
= \R^2$ and applying \refL{YudovichLemma}, it follows that
\[
    L_\nu(t) = \norm{f_\nu(t)}_{L^1(\R^2)}
        \le R \nu t + \int_0^t \beta\pr{\norm{f_\nu(s)}_{L^1(\R^2)}} \, ds,
\]
or,
\[
    L_\nu(t) \le R \nu t + \int_0^t \beta(L_\nu(s)) \, ds.
\]

If $R = 0$, Osgood's lemma immediately gives $L_\nu \equiv 0$. If $R
> 0$, we conclude that
\[
    -\Cal{M}(L_\nu(t)) + \Cal{M}(R \nu t)
        \le \int_0^t \, ds = t;
\]
that is,
\begin{align}\label{e:LtInequality}
    \int_{R \nu t}^{L_\nu(t)} \frac{ds}{\beta(s)}
        = \int_{R \nu t}^1 \frac{ds}{\beta(s)}
        - \int_{L_\nu(t)}^1 \frac{ds}{\beta(s)} \le t.
\end{align}

%
%------ Revision: 24 April 2004
%
% The proof of convergence follows a suggestion of the referee,
% though I did not define an auxiliary function that is the
% supremum of $L(t)$ over the interval $(0, t)$, because I
% didn't feel it was necessary. This is a nicer way to show
% convergence, but it necessitated moving a part of my original
% argument to the next section on the convergence rate.
%

It follows that for all $t$ in $(0, T]$,
\begin{align}\label{e:LtInequality2}
    \int_{R \nu t}^1 \frac{ds}{\beta(s)}
        \le T + \int_{L_\nu(t)}^1 \frac{ds}{\beta(s)}.
\end{align}
As $t \to 0^+$, the left side of \refE{LtInequality2} becomes
infinite; hence, so must the right side. But this implies that
$L_\nu(t) \to 0$ as $\nu \to 0^+$, and that the convergence is
uniform over $[0, T]$. $\square$

%
%------ Revision: 24 April 2004
%
% The example vorticity bounds of Yudovich that were in the first
% paragraph of this section were moved to the section Yudovich's
% Unbounded Vorticity. Also, the last sentence in the paragraph
% above was inserted.
%
% The section was also rewritten to incorporate the changes
% to the end of the proof of \refT{YudovichTheorem} in the
% previous section.
%

%
%------------ Rates of Convergence -----------------
%
\section{Rates of Convergence}\label{S:RatesOfConvergence}

\noindent Define $f \colon \R^+ \to \R^+$ implicitly by
\[
    \int_x^{f(x)} \frac{ds}{\beta(s)} = T.
\]
As $x$ decreases to zero, $f(x)$ monotonically decreases (to zero)
because $\beta$ is positive. Also, because of \refE{LtInequality},
$L_\nu(t) \le f(R \nu t) \le f(R \nu T)$, giving an expression for
a uniform bound on the convergence rate. When $1/\beta$ can be
explicitly integrated, a bound on the rate can sometimes be
determined in closed form. For the case of bounded vorticity, one
obtains essentially the same bound on the rate as in \cite{C1996}.
The sequence of bounds on vorticity in \refE{YudovichExamples} can
also be handled this way, using the upper bound on the
corresponding $\beta$ functions that Yudovich derives in
\cite{Y1995}. In the notation of \refS{YudovichBounds} this is
$\beta(x) = x \psi(1/x) \le e x \theta_{m + 1}(1/x)$.

In general, though, one can bound the initial vorticity by an
admissible function that will yield an arbitrarily slow bounded
rate of convergence. This is because the function $f$, which was
defined implicitly in terms of $\beta$, can, conversely, be used
to define $\beta$, and we can choose $f$ so that it approaches
zero arbitrarily slowly.

%
%------------ Proof of "Chemin's" Theorem -----------------
%
\section{Proof of \refT{FromCheminApproach}}\label{S:ProofOfThm23}

In this section we establish \refT{FromCheminApproach}, following
Chemin's approach in \cite{C1996}. We consider three cases: 1) both
$v_\nu$ and $v'_\nu$ are solutions to ($NS_\nu$); 2) $v_\nu$ is a
solution to ($NS_\nu$) while $v'_\nu$ is a solution to ($E$); 3) both
$v_\nu$ and $v'_\nu$ are solutions to ($E$).

Consider case 1. It follows from \refT{FactsAboutSolution} that $v_\nu$
is in $L^\iny_{loc}(\R; L^p(\R^2))$ for all $p$ such that $2 < p \le
\iny$; applying H\"{o}lder's inequality gives $v_\nu \cdot \grad v_\nu$
in $L^\iny_{loc}(\R; L^2(\R^2))$. An argument involving a Riesz
transform (as in the proof of Yudovich's theorem in \cite{C1998}, the
extra viscosity term vanishing) then shows that $\grad p_\nu$ is in
this same space.

\begin{sloppypar}
    The assumption that $\grad v_\nu$ is in $L_{loc}^\iny(\R; L^2(\R^2))$
    is enough to conclude via \refT{FactsAboutSolution} that $w_\nu$ is in
    $L_{loc}^\iny(\R; W^{1, 2}(\R^2))$ and that $\Delta v_\nu$ is in
    $L_{loc}^\iny(\R; W^{-1, 2}(\R^2))$. It then follows from ($NS_\nu$) that
    $\prt_t v_\nu$ is also in $L_{loc}^\iny(\R; W^{-1, 2}(\R^2))$. (For
    solutions to ($E$), we reach the stronger conclusion that $\prt_t
    v_\nu$ is in $L_{loc}^\iny(\R; L^2(\R^2))$.)
\end{sloppypar}

Taking the inner product of both sides of the first equation in
($NS_\nu$) with $w_\nu$ and subtracting the resulting equations for
$v_\nu$ and $v'_\nu$ gives
\begin{align}\label{e:FirstDiff}
    \begin{split}
        w_\nu \cdot \prt_t w_\nu &+ w_\nu \cdot(v_\nu \cdot \grad
                    w_\nu) \\
                &= -w_\nu \cdot \grad(p_\nu - p'_\nu) +
                    \nu w_\nu \cdot \Delta w_\nu
                    - w_\nu \cdot (w_\nu \cdot \grad v'_\nu).
    \end{split}
\end{align}

Integrating both sides of \refE{FirstDiff} over $[0, T] \times \R^2$,
the pressure term disappears because $w_\nu$ is divergence-free.
Similarly, the term $w_\nu \cdot(v_\nu \cdot \grad w_\nu)$ disappears
because $v_\nu$ is divergence-free, and we obtain
\begin{align*}
    \int_0^T \int_{\R^2} w_\nu \cdot \prt_t w_\nu  \, dx \, dt
        = \int_0^T \int_{\R^2} \nu w_\nu \cdot \Delta w_\nu
            - w_\nu \cdot (w_\nu \cdot \grad v'_\nu) \, dx \, dt.
\end{align*}

But $w_\nu$ in $L_{loc}^\iny(\R; W^{1, 2}(\R^2))$ and $\prt_t w_\nu$ in
$L_{loc}^\iny(\R; W^{-1, 2}(\R^2))$ is sufficient to conclude (see, for
instance, Lemma 1.2 p. 176 of \cite{T2001}) that
\begin{align*}
    \int_0^T \int_{\R^2} w_\nu \cdot \prt_t w_\nu  \, dx \, dt
       &= \frac{1}{2}\LTwoNorm{w_\nu(T)}^2,
\end{align*}
where we have used $w_\nu(0) = 0$.

It follows that
\begin{align}\label{e:ProtoEnergyBound1}
    \textbf{1: }\LTwoNorm{w_\nu(T)}^2
        = 2 \int_0^T \int_{\R^2} \nu  w_\nu \cdot \Delta w_\nu
            - w_\nu \cdot (w_\nu \cdot \grad v'_\nu) \, dx \, dt.
\end{align}
From the absolute continuity of the integral, we also conclude that
$\LTwoNorm{w_\nu(T)}^2$ is an absolutely continuous function of $T$.

Following a similar procedure for the other two cases, we obtain
\begin{align}
    \textbf{2: } \LTwoNorm{w_\nu(T)}^2
        &= 2 \int_0^T \int_{\R^2} \nu w_\nu \cdot \Delta v_\nu
                - w_\nu \cdot (w_\nu \cdot \grad v'_\nu) \, dx \, dt,
                \label{e:ProtoEnergyBound2} \\
    \textbf{3: }\LTwoNorm{w_\nu(T)}^2
        &= - 2 \int_0^T  \int_{\R^2} w_\nu \cdot (w_\nu \cdot \grad v'_\nu) \, dx \, dt.
                \label{e:ProtoEnergyBound3}
\end{align}

For the term common to
\refE{ProtoEnergyBound1}-\refE{ProtoEnergyBound3},
\begin{align*}  % \label{CommonBound}
    \myabs{\int_0^T \int_{\R^2} w_\nu \cdot (w_\nu \cdot \grad v'_\nu)
                \, dx \, dt}
        \le \int_0^T \int_{\R^2} \abs{w_\nu}^2 \abs{\grad v'_\nu}^2
                \, dx \, dt.
\end{align*}

Since $w(t)$ is in $W^{1, 2}(\R^2)$ for all time $t$,
\begin{align}\label{e:wLaplacianw}
    \int_0^T \int_{\R^2} w_\nu \cdot \Delta w_\nu \, dx \, dt
        = -\int_0^T \int_{\R^2} \abs{\grad w_\nu}^2 \, dx \, dt
        \le 0.
\end{align}

Similarly,
\begin{align*} % \label{e:wLaplacianw2}
    \begin{split}
        &\myabs{\int_0^T \int_{\R^2} w_\nu \cdot \Delta v_\nu \, dx \, dt} \\
            &\qquad\le \norm{\grad v_\nu}_{L^\iny([0, T];L^2(\R^2))}
                    \norm{\grad w_\nu}_{L^\iny([0, T];L^2(\R^2))} T \\
            &\qquad\le C \smallnorm{\omega^0}^2_{L^2(\R^2)} T.
    \end{split}
\end{align*}

Putting this all together gives \refT{FromCheminApproach} with, for the
three cases,
\begin{align*}
    \textbf{1: } \; R = 0, \quad
    \textbf{2: } \; R = C \smallnorm{\omega^0}^2_{L^2(\R^2)} > 0, \quad
    \textbf{3: } \; R = 0.
\end{align*}
In case 1, we only know that $R$, which comes from \refE{wLaplacianw},
is negative or equal to 0; we cannot choose, a priori, a specific
constant other than 0.

If $v_\nu$ and $v'_\nu$ were solutions for different initial
conditions, then \refE{ProtoEnergyBound1} and \refE{ProtoEnergyBound3}
would have the additional term $\LTwoNorm{\omega_\nu(0)}^2$ on the
right-hand side. Modifying the argument in Section 2 to incorporate
this term is the basis of the proof of existence in \refT{MainResult}.

%
%---------------------------- Acknowledgement -----------------------------
%
\section*{Acknowledgement}\label{S:Acknowledgement}

I wish to thank Misha Vishik for recommending that I read \cite{C1996}
and \cite{Y1995} and combine the two results.

\bibliographystyle{amsalpha}

\end{document}